\documentstyle[11pt,paspconf,epsf]{article}
\begin{document}
\pagestyle{empty}
\title{Dark Matter in Low Surface Brightness Galaxies}
\author{W.J.G. de Blok}
\affil{Kapteyn Astronomical Institute, P.O. Box 800, NL-9700 AV,
Groningen, The Netherlands}
\author{S.S. McGaugh}
\affil{Department of Terrestrial Magnetism, Carnegie Institute of Washington, 5241 Broad Branch Road NW, Washington DC 20015, USA}
\begin{abstract}
  Low Surface Brightness (LSB) galaxies form a large population of
  disc galaxies that extend the Hubble sequence towards extreme
  late-types.  They are only slowly evolving, and still in an early
  evolutionary state. The Tully-Fisher relation and rotation curves of
  LSB galaxies both show that LSB galaxies are very dark-matter
  dominated with respect to ``normal'' high surface brightness (HSB)
  galaxies. Mass models derived from the rotation curves of LSB and
  HSB galaxies show that LSB galaxies inhabit less dense and more
  extended halos. Mass density, which changes with surface brightness, 
  is as important in determining the evolution of a galaxy as total
  mass is.
\end{abstract}
\section{Introduction}

Increasingly sensitive surveys of our local universe have in recent
years revealed the existence of a population of dim field galaxies
(e.g.\ Schombert et al.\ 1992).  These galaxies are primarily
late-type disc galaxies with central surface brightnesses $\mu_0(B)
\sim 23.5$ mag arcsec$^{-2}$ and scale lengths of a few kpc.  The existence of such a population is
not consistent with the surface brightness distribution for disc
galaxies as derived by Freeman (1970), which gives a narrow range of
$\mu_0(B) = 21.65 \pm 0.30$ $B$-mag arcsec$^{-2}$. 
A better understanding of
selection effects (Disney 1976, McGaugh 1996) and deeper surveys have
shown that for every magnitude one goes fainter in surface brightness
approximately equal numbers of galaxies are found (McGaugh 1996).

Our knowledge and views of the local galaxy population have therefore
been biased towards the bright HSB galaxies.  As LSB galaxies {\it
  are} numerically important and can contain an appreciable fraction
of the total mass in galaxies (McGaugh 1996), studying them is of great
importance for improving our understanding of the local galaxy
population.

In the above I have implicitly distinguished between HSB and LSB
galaxies as if they were two separate kinds of galaxies.  But LSB
galaxies are in many ways a continuation of the classical Hubble
sequence towards extreme late-types, and any distinction on the basis
of surface brightness is artificial. Recent investigations using
well-defined samples (de Jong 1995) show that disc (spiral)
galaxies exhibit a continuous range in central surface brightness,
from the Freeman value, which really seems to be an upper limit to the
surface brightness a galaxy can attain, to the faintest values
currently detectable (and we do not seem to have reached the faint
limit yet).

For the sake of convenience I will retain the HSB/LSB distinction,
with where I will call ``HSB'' those galaxies with $\mu_0(B) \la 23$
mag arcsec$^{-2}$, and ``LSB'' those galaxies with $\mu_0(B)$ fainter
than this value. When comparing HSB and LSB galaxies it will be
implicitly assumed that galaxies of similar luminosity/rotation
velocity  are compared.

\section{General properties}

One (by now superseded) way to explain the low surface brightnesses of LSB
galaxies was the so-called ``faded-disc scenario''.  In this picture
LSB galaxies are simply ordinary HSB galaxies that for some reason
(e.g.\ gas depletion) stopped forming stars a few Gyr ago, and are now
simply fading away.  LSB galaxies would thus have to be much redder
than ``normal'' galaxies, and presumably more gas-poor.

Measurements of the colours of LSB galaxies immediately showed that this
scenario could not be the correct explanation.  McGaugh (1992), van der
Hulst et al.  (1993), McGaugh \& Bothun (1994), de Blok et al.  (1995)
all showed that colours of LSB galaxies are among the bluest known for
non-interacting disc galaxies.

Two possible causes for these blue colours are low metallicity and
recent star formation.  Both are found in LSB galaxies. H$\alpha$
imaging (McGaugh 1992) shows that a few regions of star formation are
usually present in LSB galaxies.  The low surface brightness of the
underlying disc ensures that only a small amount of star formation is
needed to significantly influence the colours (de Blok et al. 1995).
Measurements of the oxygen abundances in H{\sc ii} regions in
LSB galaxies (McGaugh 1994) show that the metallicity is on
average 0.2-0.5 solar.

LSB galaxies in addition contain significant amounts of H{\sc i}
($\sim 10^9 M_{\odot}$) (van der
Hulst et al. 1987; Schombert et al. 1992; McGaugh 1992).  The
low metallicity, normal H{\sc i} masses and the small amounts of star
formation all suggest slow evolution.  This is supported by
measurements of the H{\sc i} surface densities. Van der Hulst et al.
(1993) showed that the H{\sc i} surface densities are significantly
lower than in late type HSB galaxies.  This was interpreted using the
critical star formation treshold (Kennicutt 1989) as a possible cause
for the low star formation rate in LSB galaxies.  Follow-up
observations of a larger sample of more extreme LSB galaxies (de Blok
et al. 1996) confirmed the low H{\sc i} surface densities, but also
showed that the extended H{\sc i} discs still contain enough gas
to make LSB galaxies rank among the most H{\sc i}-rich disc galaxies.

The unevolved nature of LSB galaxies is also supported by modelling of
their spectro-photometric evolution (van den Hoek et al.\ 1996).  The
current star formation rates are found to be only $\sim 0.1$
$M_{\odot}{\rm yr}^{-1}$, which is an order of magnitude lower than in
HSB Sc galaxies, where values of $\sim 4$ $M_{\odot}{\rm yr}^{-1}$ are
usually found (Kennicutt 1983).  The gas depletion time scales in LSB
galaxies are thus much longer.  A typical HSB galaxy will exhaust its
gas reservoir in a few Gyr, while in a LSB galaxy timescales of
tens of Gyr are inferred.

As the evolution time scales are thus much longer, and if the
evolution processes are similar, this essentially implies that LSB
galaxies are just HSB galaxies in the making, but at a very slow rate.
Or, in other words, LSB galaxies give us the chance to study the
conditions as they probably were in late-type HSB galaxies of
comparable mass many Gyr ago.

\section{LSB galaxies: Nature or Nurture?}

But what makes LSB galaxies LSB? It cannot be a lack of
``evolutionary fuel'': they still posses large amounts of hydrogen. 

Can it be environment? Bothun et al.\ (1993) and Mo et al.\ (1995)
found that, although LSB galaxies do follow the general large scale
structure delineated by HSB galaxies, they tend to be more isolated at
scales $<$ 2 Mpc.  LSB galaxies thus avoid other galaxies and
furthermore show no evidence of having companions (Bothun et al.
1993).  A lack of interactions might thus have produced this quiescent
evolution. The relatively high star formation rates, grand design
spirals and other features that make HSB galaxies so conspicuous would
thus have to be enhanced partly by interactions and tidal effects.
This would also explain why LSB disc galaxies are preferably found in the
field or in the outskirts of clusters: if they fall deeper into the
cluster potential, interactions with the cluster galaxies or
intra-cluster medium will will either enhance their star formation
rates, raising their surface brightnesses or strip them of their large,
low density gas discs, thus eventually reddening their colours.  In
both cases we would no longer identify them as proto-typical LSB
galaxies.

Or could it be that the dark halos regulate the evolution of their
inhabitants? Persic \& Salluci (1991) pointed out that the importance of
dark matter (DM) with respect to the luminous matter increases towards
late Hubble types.  As LSB galaxies do extend the Hubble sequence
towards later types, one would expect the DM to be very important.
Could the dominant DM thus stabilize the baryonic disc
against the instabilities that would normally lead to star
formation? 

If both scenarios are tthe darkest galaxies
are always the most isolated from their neighbours? It would imply
that the eventual LSB or HSB fate is already determined at the time of
galaxy formation.

A first clue about the importance of DM can be derived from the
Tully-Fisher (TF) relation.  Zwaan et al.\ (1995) and Sprayberry et
al.\  (1995) have shown that LSB galaxies do follow the exactly the
same TF relation as HSB galaxies. This is not something one would
expect {\it a priori}.  The TF relation can be written as
$L[\Sigma_0(M/L)^2]\propto V^4$,
where $M/L$ is measured within a fixed number of scale lengths, and
$\Sigma_0$ is the central surface brightness $\mu_0$ expressed in linear units.

\begin{figure}
\plottwo{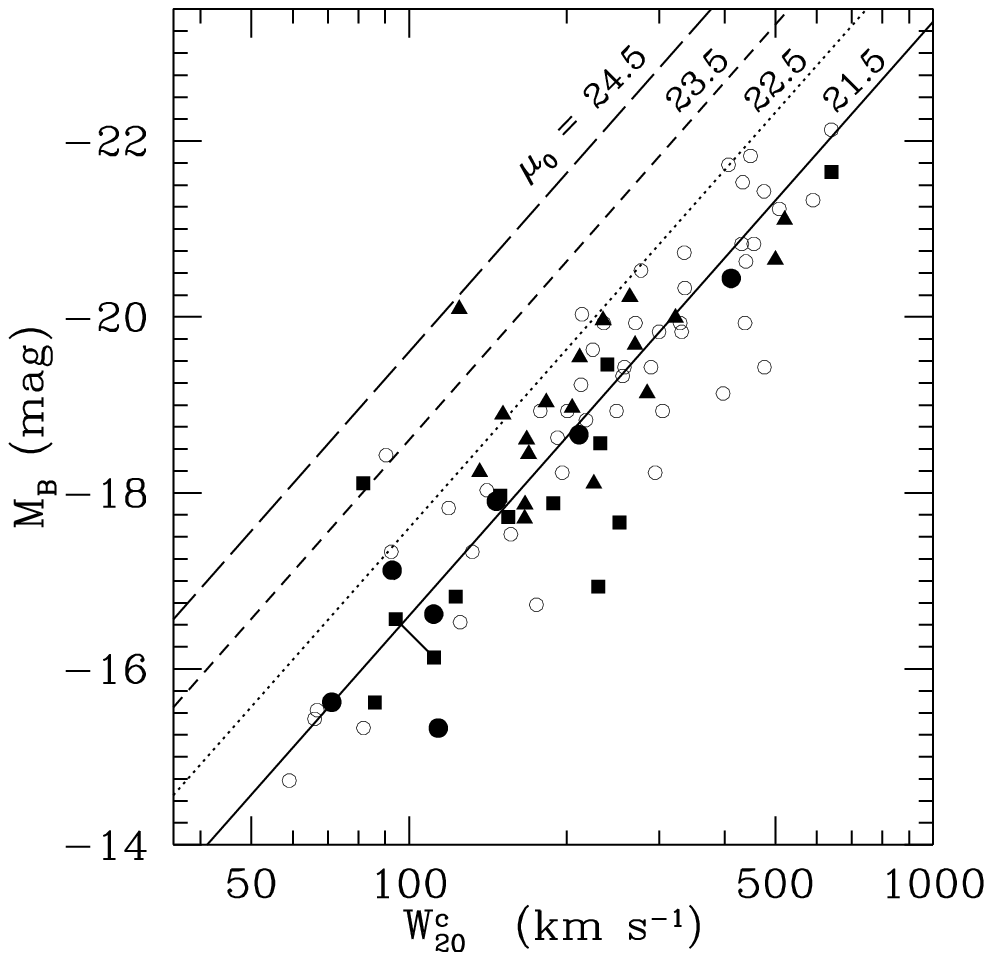}{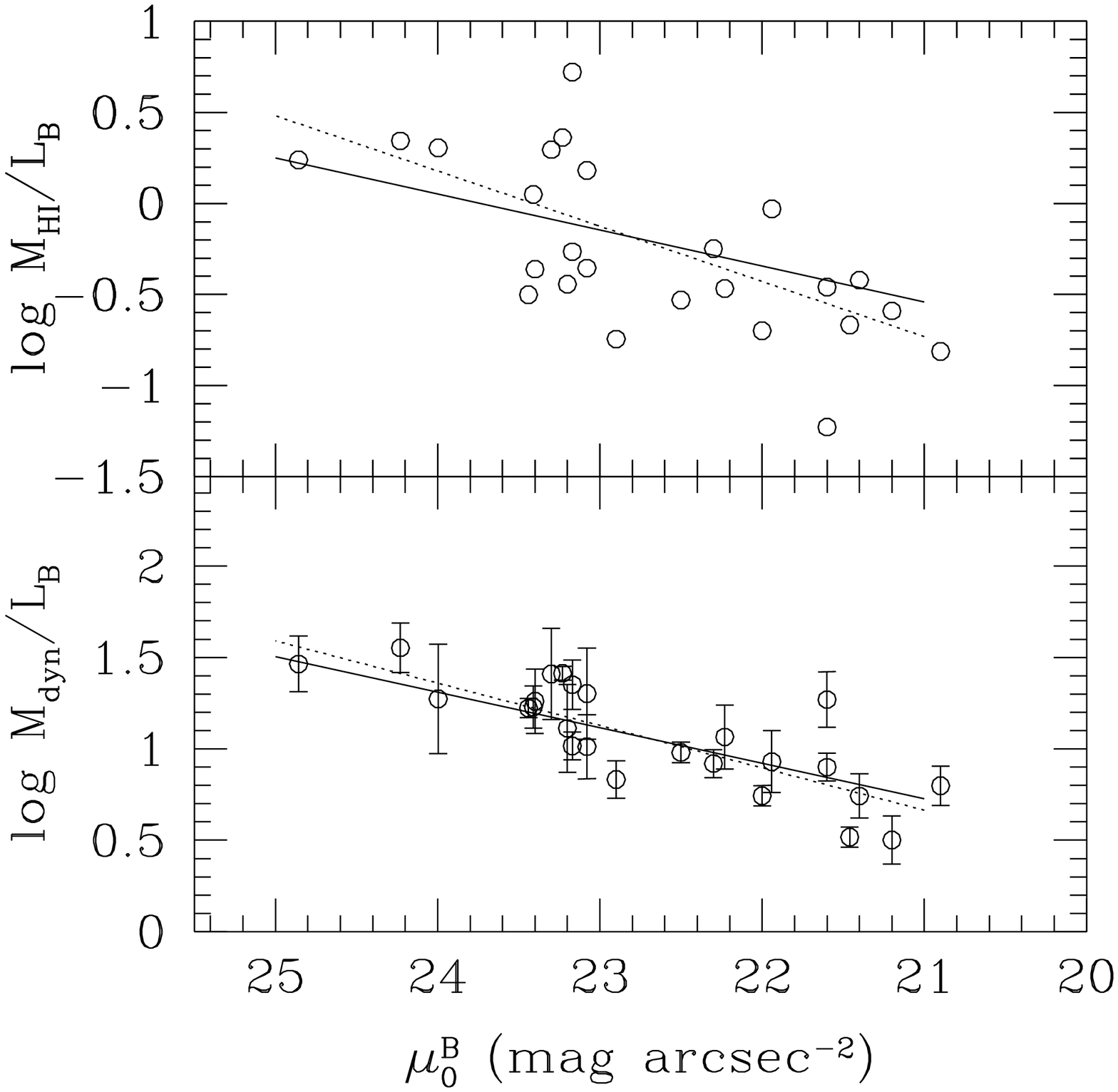}
\caption{Left: TF relation for HSB and LSB galaxies (Zwaan et al. 1995). Open
  circles are the comparison sample of Tully \& Fouqu\' e (1985).
  Triangles are LSB galaxies with 22 $< \mu_0 <$ 23; squares indicate
  23 $\le \mu_0 <$ 24; filled circles are galaxies with $\mu_0 \ge
  24$. Diagonal lines are TF relations for the indicated values of
  $\mu_0$ if all galaxies had identical $M/L$.  Right: $M_{\rm
    HI}/L_B$ and $M_{\rm dyn}/L_B$ within 4$h$ versus $\mu_0$. The
  samples of Broeils (1992) and de Blok et al. (1996) are shown.
  Dotted lines as least-squares fits; full lines are predictions from
  the TF relation.  Errorbars are based on an inclination uncertainty 
  of 6$^{\circ}$.}
\end{figure}

If all galaxies had similar total $M/L$ ratios, one would expect
parallel TF relations for each surface brightness, as is shown in Fig
1.  As this is evidently not the case, there must exist a tight
coupling between the surface brightness of a galaxy and its
$M/L$ ratio, which implies that LSB galaxies
must have higher $M/L$ values than HSB galaxies.
This also means that they are more DM dominated (see below).

The relation between $\mu_0$ and $M/L$ is actually observed (Fig. 2).
Also shown is the relation between $\mu_0$ and $M_{\rm HI}/L$.  Apart
from the fact that it clearly shows that there is indeed a systematic
trend with surface brightness, the identical slopes of both trends also
suggest a link between evolution rate and DM dominance.

\section{Mass models of LSB galaxies}

To investigate the systematics between the DM fraction and surface
brightness de Blok et al.\ (1996) obtained rotation curves of a sample
of 19 LSB galaxies, and  compared these to a large sample of
curves of HSB galaxies (de Blok \& McGaugh 1996a, 1996b, Broeils 1992).
In Fig. 2 an example of the LSB galaxy data is shown.
\begin{figure}
\plotone{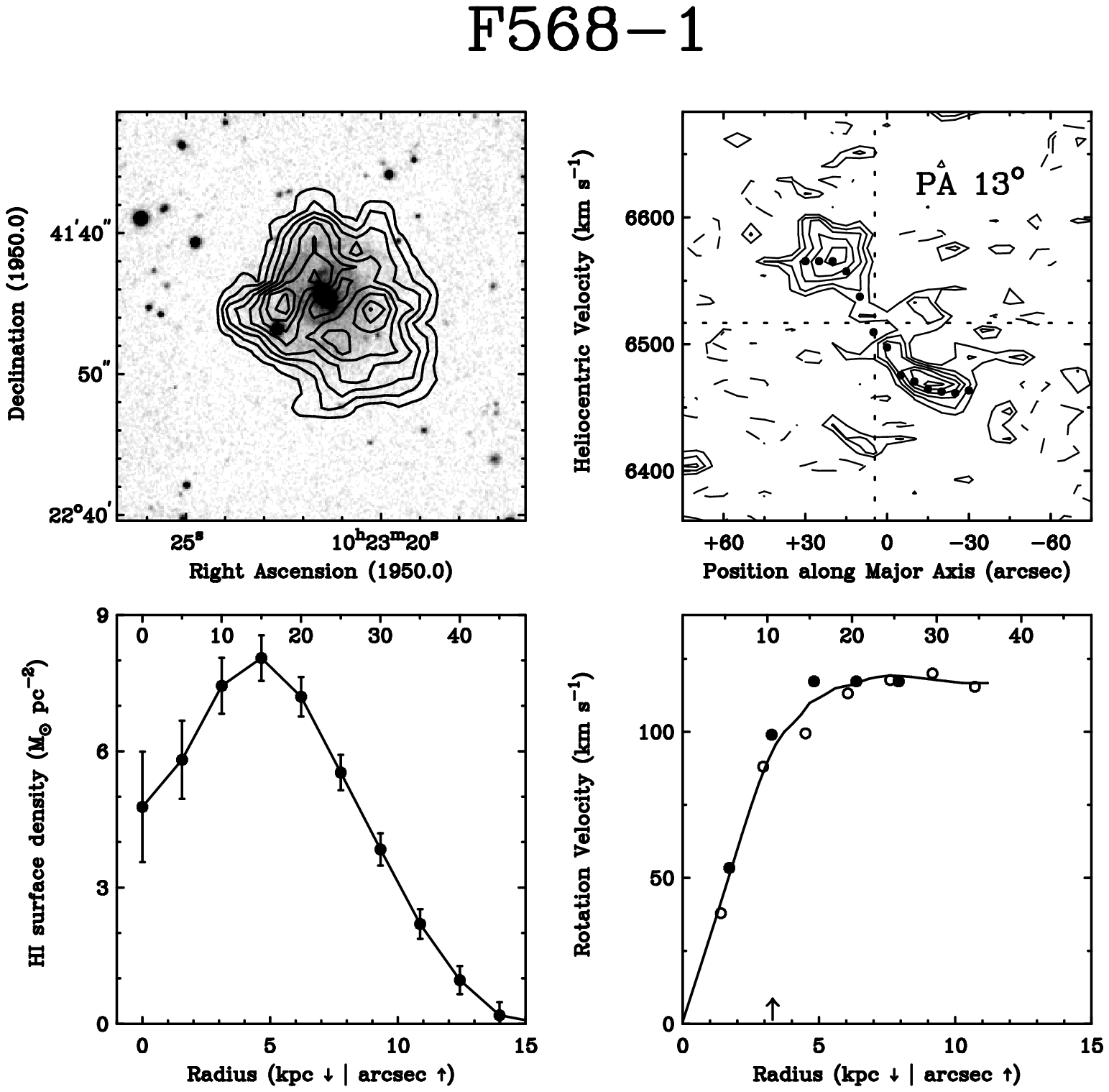}
\caption{LSB galaxy F568-1 (de Blok et
  al. 1996). Top-left: $R$-band image with
  HI surface density contours. Contour levels:
  1,3,5...$\times 10^{20}$ cm$^{-2}$. Top-right: 
  major axis position velocity diagram. Contours:
  $-2,2,3,4...\sigma$. Bottom-left: inclination
  corrected radial surface density profile. Bottom-right:
  inclination corrected rotation curve. Open circles
  denote approaching side, filled circles receding
  side.}
\end{figure}
The rotation curves of LSB galaxies are only slowly rising, and
usually continue to do so until the outermost measured points. A few
curves do show signs of flattening off at the outermost radii.
Maximum velocities are typically between 50 and 120 km s$^{-1}$.  The
LSB galaxies differ from the HSB galaxies in that at fixed $V_{\rm
  max}$ they have larger scale lengths and their rotation curves
always rise more slowly. At fixed $V_{\rm max}$ (fixed $L$) the shape
of the rotation curves thus changes with surface brightness, or,
alternatively, at fixed $h$ maximum rotation velocity decreases with
surface brightness.
\begin{table}
\begin{center}
\caption{Comparison of LSB galaxy U128 and HSB galaxy N2403}
\begin{tabular}{lll}
\tableline
Name & UGC 128 & NGC 2403 \\
\tableline
$M_B$ (mag)     & \llap{--}18.9        & \llap{--}19.2 \\
$V_{\rm max}$ (km s$^{-1}$) & 131 & 136 \\
$\mu_{0,B}$ (mag arcsec$^{-2}$)& 24.2 & 21.4\\
$h$ (kpc) & 6.8 & 2.1\\
\tableline
$R<6.2h$&&\\
\tableline
$M_{\rm gas}$ ($\times 10^{10}$ M$_{\odot}$) & $1.05 $& $0.34$\\
$M_{\rm star}^{\rm max}$ ($\times 10^{10}$ M$_{\odot}$) & $1.68 $&$1.36$\\
$M_{\rm dark}$ ($\times 10^{10}$  M$_{\odot}$) & $14.43$& $3.90$\\
$(M/L)_B$ & 30.6 &  7.4\\
$\langle \rho \rangle$\ $(\times 10^{-3} M_{\odot}{\rm pc}^{-3})$ & 2.3 & 25\\
\tableline
\tableline
\end{tabular}
\end{center}
\end{table}
A good way to quantify the differences between LSB and HSB galaxies is
to compare the properties of two galaxies at identical positions on
the TF, in order to cancel out their global differences. Two
such galaxies are NGC 2403 and UGC 128. Both have $M_B\simeq -19$ and
$V_{\rm max} \simeq 134$ km s$^{-1}$ (see Table 1).  De Blok \& McGaugh
(1996a) have made mass models of the  stellar, gas and dark components
of both galaxies, assuming for the stellar disc a maximum disc model
and an isothermal halo. In Table 1 the relevant
masses and mass-to-light ratios within 6.2$h$ are listed.  It is
clear that the $M/L$ ratios in the LSB galaxy are much
larger, and that DM is much more dominant. This conclusion remains
true even if we measure the properties at a fixed number of kpc. Note
that non-maximum disc models will only enhance the differences. UGC
128 is also more gas-rich than NGC 2403.  Furthermore, the average
density of matter within 6.2$h$ is a factor of 10 less in UGC 128 than
in NGC 2403. LSB galaxies thus seem to be low density objects.
\begin{figure}
\plotone{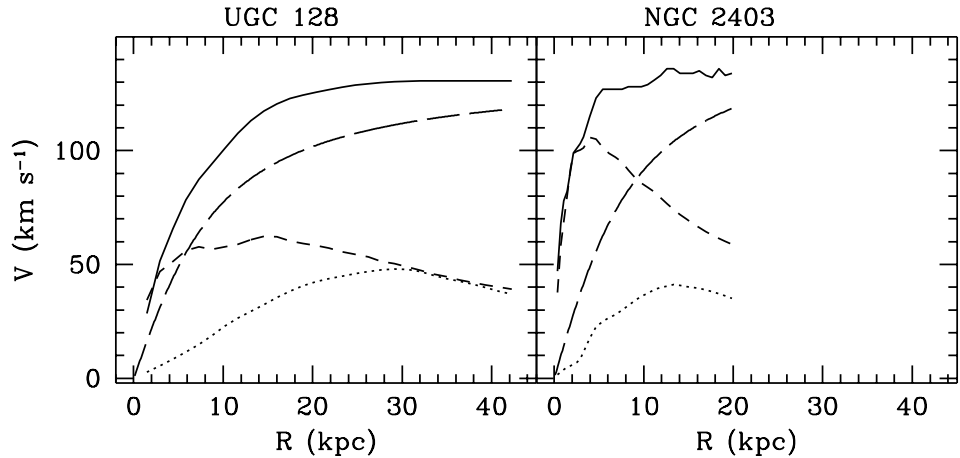}
\caption{Comparison of the rotation curves of LSB galaxy UGC 128 and
  HSB galaxy NGC 2403. The full line is the observed rotation curve,
  the dotted curve that of the H{\sc i} (scaled by 1.33 to take helium
  into account), the short-dashed line is the rotation curve of the
  maximum disc; the long-dashed
  line is the rotation curve of the halo.}
\end{figure}
This dramatic increase in the importance of DM towards lower surface
brightnesses is also found other LSB galaxies (de Blok \& McGaugh 1996b).

We have seen that LSB galaxies usually have larger optical scale
lengths and more extended gas discs. Does this also mean that their
halos are extended and low density?  An extensive discussion is given
in de Blok \& McGaugh (1996b), of which we will summarize the main
points here. The answer depends crucially on the assumed
$(M/L)_{\star}$ of the stellar disc. In making mass models of galaxies
the maximum disc hypothesis (van Albada \& Sancisi 1986) is usually
assumed: the rotation curve of the stellar disc is scaled in order to
attribute as much rotation velocity to it as possible within the
constraints of the observed rotation curve.

Applying this hypothesis to rotation curves of LSB galaxies yields
contradictory conclusions. At a fixed $V_{\rm max}$ LSB galaxies are
then always found to have the highest values of $(M/L)_{\star}$.  This
implies that LSB galaxies have very evolved discs, which is a
possibility we can exclude in the face of all other evidence (colours,
metallicities etc.; see Section 2).  Secondly, as the rotation curve
of UGC 128 in Fig.\ 3 shows, applying the maximum disc recipe has a
different result in LSB galaxies than in HSB galaxies: in LSB galaxies
it is not possible to explain all of the rotation velocity in the
inner part by just scaling the stellar disc, in contrast to HSB
galaxies.  Thirdly, it can be shown that the large $(M/L)_{\star}$
values derived for LSB galaxies are a mathematical effect of the
extendedness of their discs. They do not need to have
consequences for conclusions on their evolutionary state.
Maximum disc is therefore not the preferred way of making mass models
of LSB galaxies.  Instead $(M/L)_{\star}$ ratios derived from stellar
velocity dispersions and colours [the ``Bottema'' disc, see Bottema
(1995, 1996) and de Blok \& McGaugh (1996b)] are used. This appears nearly
equivalent to a constant $(M/L)_{\star}$ assumption. For HSB galaxies
a typical value of $(M/L)_{\star} \simeq 1.8$ is found, while for LSB
galaxies $(M/L)_{\star} \simeq 1.4$.
Mass models of a sample of HSB and LSB galaxies derived assuming the
Bottema disc show that LSB galaxies tend to have lower values of the
central halo density $\rho_0$ and higher values of the core radius
$R_C$ than HSB galaxies (Fig.~4). This conclusion does of course
depend on the validity of the Bottema disk. However, Bottema (1995)
has convincingly shown that the Bottema disk is a good working
hypothesis in HSB galaxies, while in LSB galaxies 
the DM dominance
makes the halo parameters relatively insensitive to $(M/L)_{\star}$
assumptions.


\begin{figure}
\plotone{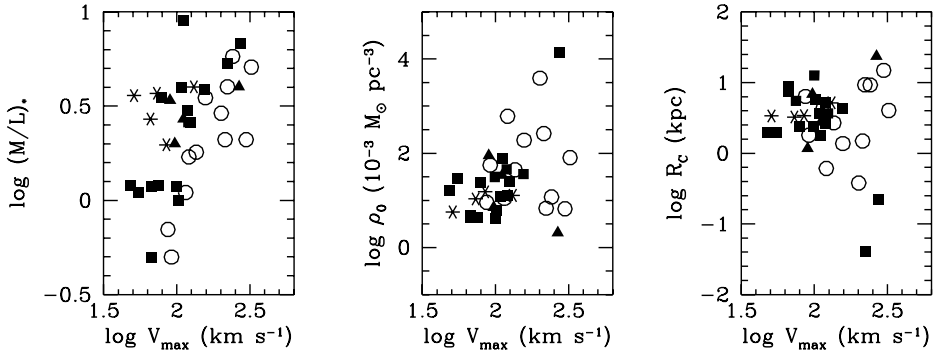}
\caption{Maximum disc $(M/L_B)_{\star}$ ratios (left), 
 central halo density $\rho_0$ (center) and core radius $R_C$ (right)
  (both derived assuming Bottema disc) 
  of LSB and HSB galaxies plotted versus observed maximum rotation
  velocity. Open circles indate galaxies with $\mu_0 < 22$; triangles
  $22 \le \mu_0 < 23$; squares $23 \le \mu_0 < 24$; asterisks $\mu_0
  \ge 24$.}
\end{figure}

\section{Conclusions}

LSB galaxies are extended, low density galaxies, that are still in an
early stage of galaxy evolution. They inhabit less dense and more
extended halos than HSB galaxies. Colours are blue, metallicities are
low, they have high values of $M_{HI}/L_B$ and only small amounts of
star formation.  Evolutionary time scales are in general much longer
than in HSB galaxies.  The fact that LSB galaxies do have luminosities
and masses comparable to those of HSB galaxies, evolve much slower,
and are found in low density regions of the general galaxy
distribution, does show that galaxy evolution does not just depend on
mass (observed as luminosity), but just as much on density (observed
as surface brightness) and the environment where a galaxy is formed.


\begin{references} 
\reference Bothun, G.D., Schombert, J.M.,  Impey, C.D., Sprayberry,
D.,  McGaugh, S.S. 1993, \aj , 106, 530
\reference Bottema, R. 1995, PhD thesis, University of Groningen
\reference Bottema, R. 1996, in preparation
\reference Broeils, A.H. 1992, PhD thesis, University of Groningen
\reference de Blok, W.J.G., van der Hulst, J.M., Bothun, G.D. 1995,
\mnras , 274, 235
\reference de Blok, W.J.G., McGaugh, S.S. 1996a, ApJL, 469, L89
\reference de Blok, W.J.G., McGaugh, S.S. 1996b, \mnras , submitted
\reference de Jong, R.S. 1995, PhD thesis, University of Groningen
\reference Disney, M.J. 1976, Nature, 263, 573
\reference Freeman, K.C. 1970, \apj , 160, 811
\reference Kennicutt, R.C., Jr. 1983, \apj , 272, 54
\reference Kennicutt, R.C., Jr. 1989, \apj , 344, 685
\reference McGaugh, S.S. 1992, PhD thesis, University of Michigan
\reference McGaugh, S.S. 1994, \apj , 426, 135
\reference McGaugh, S.S., Bothun, G.D. 1994, \aj , 107, 530
\reference McGaugh, S.S. 1996, \mnras , 280, 337
\reference Mo, H.J., McGaugh, S.S., Bothun, G.D. 1994, \mnras , 267, 129
\reference Persic, M., Salucci, P. 1991, \apj , 368, 60
\reference Schombert, J.M., Bothun, G.D., Schneider, S.E.,  McGaugh,
S.S 1992, \aj , 103, 1107
\reference Sprayberry, D., Bernstein, G.M., Impey, C.D., Bothun, G.D.
1995, \apj , 438, 72
\reference Tully R.B., Fouqu\' e, P. 1985, \apjs , 58, 67
\reference van Albada, T.S, Sancisi, R. 1986,
Phil. Trans. R. Soc. Lond. A,  320, 447
\reference van den Hoek, L.B., de Blok, W.J.G., van der Hulst, J.M.,
de Jong, T. 1996, submitted
\reference van der Hulst, J.M., Skillman, E.D., Kennicutt, R.C.,
Bothun, G.D. 1987, \aap , 177, 63
\reference van der Hulst, J.M., Skillman, E.D., Smith, T.R., Bothun,
G.D., McGaugh, S.S.,  de Blok, W.J.G. 1993, \aj , { 106}, 548 
\reference Zwaan, M.A., van der Hulst, J.M., de Blok, W.J.G., McGaugh
S.S. 1995, \mnras , 273, L35
\end{references}
\end{document}